# Characterization of a Large-Area HRPPD MCP-PMT Coupled to LYSO for TOF-PET

Yingjie Wang, Mark A. Popecki, Heejong Kim, Alexey V. Lyashenko, Stefan Cwik, Mythra Varun Nemallapudi, Jeffrey S. Souris, Michael J. Minot, Chien-Min Kao, and Chin-Tu Chen

***Abstract*—Large-area microchannel-plate photomultipliers (MCP-PMTs) such as the High-Rate Picosecond Photon Detector (HRPPD) are attractive for time-of-flight positron emission tomography (TOF-PET). They offer excellent single-photon time resolution, low dark count rate, and large monolithic active areas. However, the coincidence time resolution (CTR) reported so far with PET scintillators falls short of expectations based upon their single-photon time resolution. Here we report a characterization of a 104×104 mm² HRPPD (active area) coupled to LYSO crystals and read out through a custom high-frequency amplifier. With a 2×2×3 mm³ LYSO crystal we measured a CTR of 107 ps full width at half maximum (FWHM) against a silicon photomultiplier (SiPM)-based reference detector. After subtracting the independently calibrated reference detector's contribution of 42 ps, the single HRPPD detector was found to contribute 98 ps, corresponding to 139 ps for two identical HRPPD detectors operating in coincidence mode. With clinical 3×3×20 mm³ crystals the CTR against the reference SiPM detector was 160 ps, corresponding to 218 ps for two identical HRPPD detectors. Timing and gain were uniform across output pads within measurement precision. Summing the four pads under a 3×3×10 mm³ LYSO crystal, the energy resolution for 511 keV gamma ray was 14.6% FWHM, and the 1275 keV line of the $^{22}$Na source was within 1% of its proportional position relative to the 511 keV positron annihilation peak. Flood maps resolved the individual 3 mm elements of a 4×4 array despite the 25 mm pad pitch. To the best of our knowledge the observed timing resolution improves substantially on what has been reported for large-area MCP-PMTs coupled to PET scintillators, and the crystal array identification has not been demonstrated before on such a device. These results indicate that this device class deserves renewed consideration as a photodetector technology for high-performance TOF-PET systems.***

This work was supported in part by the Ariadne Beck Trust Fund. *(Corresponding author: Yingjie Wang.)* Michael J. Minot, Chien-Min Kao, and Chin-Tu Chen equally supervised this work.

Yingjie Wang, Heejong Kim, Mythra Varun Nemallapudi, Jeffrey S. Souris, Chien-Min Kao, and Chin-Tu Chen are with the Department of Radiology, University of Chicago, Chicago, IL 60637 USA (e-mail: wangyingjie@uchicago.edu).

Mark A. Popecki, Alexey V. Lyashenko, Stefan Cwik, and Michael J. Minot are with Incom Inc., Charlton, MA 01507 USA.

## I. Introduction

In the time-of-flight positron emission tomography (TOF-PET), the signal-to-noise ratio (SNR) of the reconstructed image scales inversely with the square root of the coincidence time resolution (CTR; quoted as full width at half maximum, FWHM, throughout this paper). Hence a better CTR would allow shorter scans, lower injected dose to a patient, or imaging at lower count rates. For CTRs below ~40 ps, the photon arrival time measurements alone constrain annihilations to within 6 mm along the line of response, approaching reconstruction-less imaging. Instrumentation improvements over the past decades have improved the CTRs of clinical systems 3-fold, from ~600 ps to ~200 ps [1], [2]. Further enhancement of the CTR is one of the major technological challenges faced by PET systems developers.

Achieving such timing resolution improvements places stringent demands on photodetector design, calling for high detection efficiency, excellent single-photon time resolution (SPTR), high gain, and low noise. Microchannel-plate photomultiplier tubes (MCP-PMTs) have long been among the best candidates for fast-timing photon detection, owing to their high gain ($10^6$–$10^8$), superior SPTR (10–20 ps $\sigma$), and low dark count rate [3]. To date, Incom Inc. has developed two large-active-area MCP-PMTs, the 193×193 mm² Large-Area Picosecond Photon Detector (LAPPD) and the 104×104 mm² High-Rate Picosecond Photon Detector (HRPPD), both of which meet these specifications and offer flexible readout configurations [4], [5], [6], [7], [8], [9], [10], [11].

The combination of excellent SPTR and low dark count rate of MCP-PMTs has been exploited most directly in Cherenkov-based PET detectors, where the timing is carried by the few prompt photons produced when a 511 keV gamma interacts. Because that photon yield is small, these detectors commonly use an integrated window design, in which the entrance window of the tube also serves as the Cherenkov radiator to avoid extra light loss at a coupling interface. Empirical studies and simulations have shown the advantages of this approach [1], [12], [13].

The same properties are equally useful with traditional scintillation detectors. Coupling an MCP-PMT directly to a scintillator with high stopping power and fast light output, such as lutetium–yttrium oxyorthosilicate (LYSO), could still improve on state-of-the-art PET performance. Several research groups have explored LAPPD and HRPPD coupled to conventional scintillators for TOF-PET [14], [15], [16]. The best CTR reported to date for two identical detectors is about 234 ps, obtained with a 3×3×3 mm³ LYSO crystal [16], which suggests that the intrinsic SPTR of these devices is not yet fully exploited. To the best of our knowledge, no prior work has reported resolved crystal arrays coupled to a LAPPD or HRPPD.

The output signal from an MCP-PMT rises in under a nanosecond, so exploiting that speed requires an amplifier matched to the signal in bandwidth and noise, with an operating point tuned over the microchannel plate (MCP) and photocathode voltages together. We therefore built a custom high frequency amplifier for HRPPD readout, applied leading edge discrimination to fully sampled waveforms, and tuned the operating point systematically. This paper reports the resulting characterization of the HRPPD coupled to LYSO, including the operating point, coincidence timing and its deconvolution against a silicon photomultiplier (SiPM)-based reference detector, crystal size impact (from the 2×2×3 mm³ benchmark to the clinical 3×3×20 mm³ geometry), pad-to-pad uniformity, 511 keV photon energy resolution and linearity, and crystal array identification.

## II. Materials and Methods

### A. HRPPD detector

The HRPPD (Incom Inc.) is a compact, sealed microchannel plate photomultiplier developed for high rate picosecond timing, with an active area of 104×104 mm², a 5 mm thick fused silica entrance window, a bialkali photocathode, a chevron pair of microchannel plates, and a segmented anode read out through a 4×4 array of capacitively coupled pads at a 25 mm pitch. The window side and the back of the device are shown in Fig. 1(a) and (b), and its layer structure in Fig. 1(d). Its construction and performance have been described by Incom and collaborators [4], [17]. Recent units reach a peak quantum efficiency above 30% and an SPTR of 15 to 20 ps σ, with a dark count rate below 1.8 kHz/cm² (at ≥ $10^6$ gain) and an afterpulse fraction below 2.8% at a gain of $5\times10^6$ and a 100 V photocathode gap voltage [18]. In operation, the device is biased through five high-voltage (HV) electrodes, HV #1 to HV #5 in Fig. 1(d), all held at negative voltage with respect to ground. The photocathode gap voltage ($V_{PG}$) is the difference between HV #1 and HV #2, and the per-MCP voltage ($V_{MCP}$) is the difference across each MCP, between HV #2 and HV #3 and between HV #4 and HV #5. Both were scanned in this study to establish the optimal operating point (Section III-A). The pads are numbered 1 to 16 as shown in Fig. 1(c), with a single output electrode brought out at the center of each pad. All measurements reported here used the lower left 2×2 subarray, pads 9, 10, 13, and 14. The capacitively coupled interface board shown in Fig. 1(b) is external to the sealed device, is provided as standard with the HRPPD, and can be removed and exchanged for one customized to meet the needs of different applications.

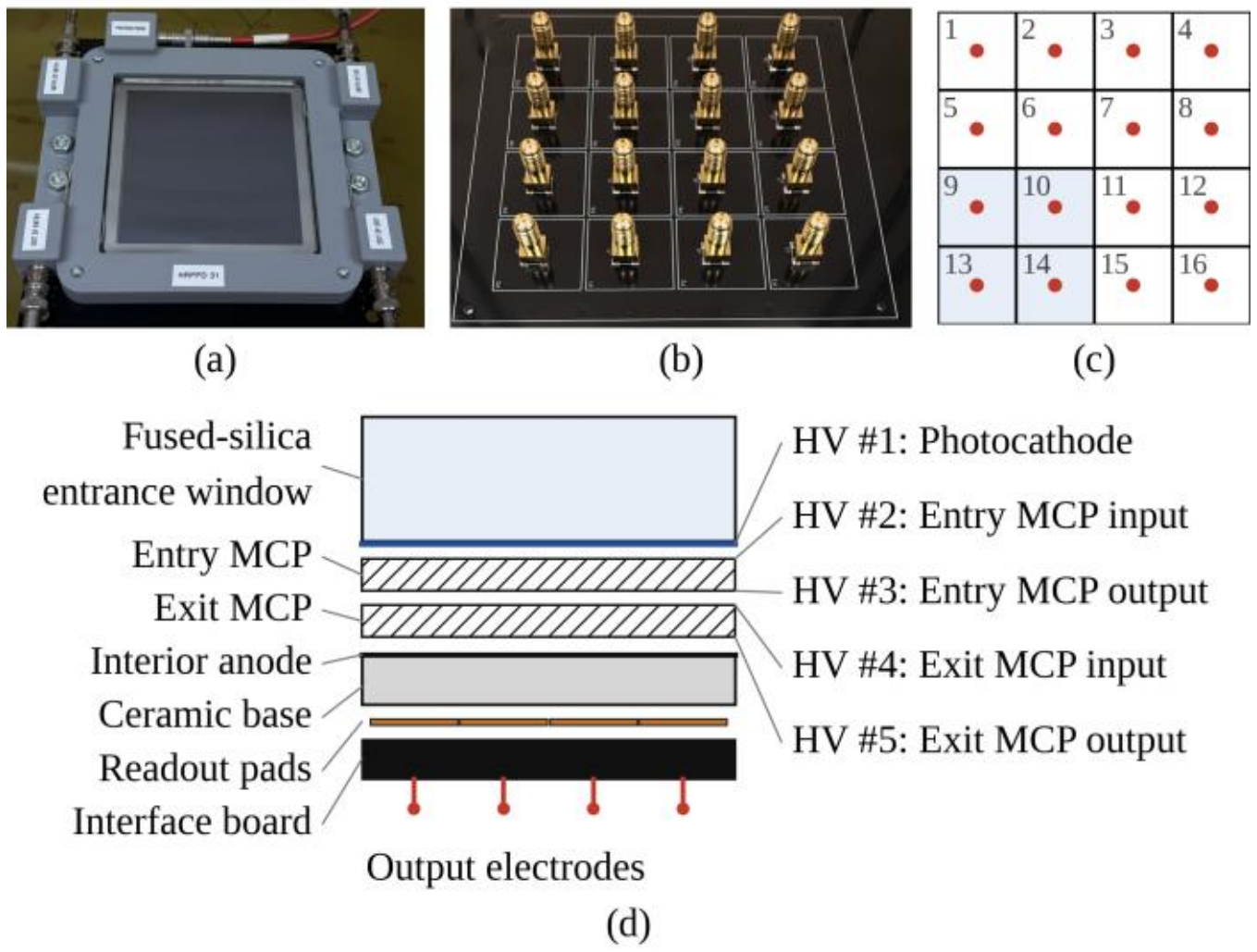

**Fig. 1.** The HRPPD used in this work. (a) Window side, with the 104×104 mm² fused-silica entrance window. (b) Back side, showing the interface board that brings out the 4×4 readout channels. (c) Pad numbering, viewed from the window side, with the output electrode of each pad marked in red. The shaded 2×2 group, pads 9, 10, 13 and 14, denotes the quadrant used for all four-channel measurements in this work. (d) Schematic cross section of the HRPPD (not drawn to scale), showing its layer structure (left) and the five high-voltage electrodes HV #1 to HV #5 (right).

### B. Readout Electronics

Custom readout electronics split the signal from a pad on the capacitively coupled readout board into separate energy and timing channels (Fig. 2). The energy channel used an AD8000 amplifier, while the timing channel used a two-stage RF amplifier (ADL5611) whose high gain and slew rate preserved the output signal's fast, large amplitude rising edge. The ADL5611's bandwidth of up to 6 GHz matches the rise time of the HRPPD signal from the LYSO crystal and leaves headroom for future Cherenkov photon related tests. The coincidence timing was recorded using a LeCroy WaveMaster 8060HD oscilloscope operated at 2 GHz analog bandwidth and a sampling rate of 20 GS/s with all four channels in use. Timing and gain uniformity were assessed by repeating the same measurement on each of the four pads. The array flood maps and the energy resolution were taken instead with a DRS4 evaluation board (Paul Scherrer Institute, 5 GS/s, 700 MHz analog bandwidth), which provides the four simultaneous channels those two measurements need, with pad outputs on 0.5 m coaxial cables to minimize noise pickup.

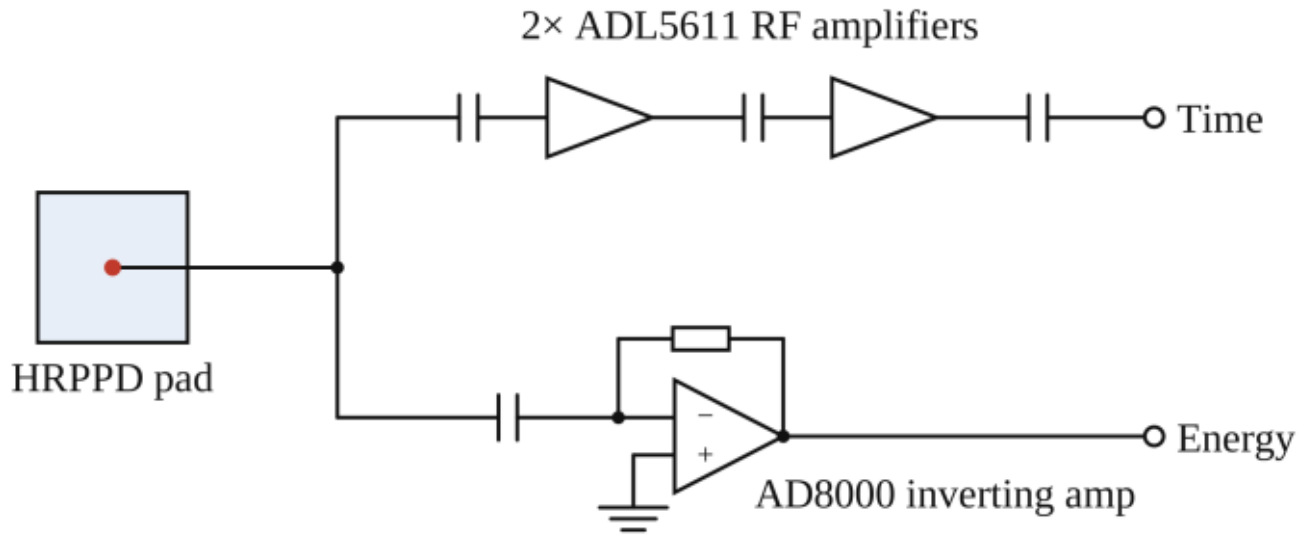

**Fig. 2.** Schematic of the custom high-frequency board connected to one HRPPD output channel. The signal is divided into a timing branch, with two ADL5611 RF amplifier stages, and an energy branch with a single AD8000 inverting amplifier.

## C. Reference SiPM Detector

The 5 mm thick entrance window spreads the light over several HRPPD channels, so a coincidence between two channels of the same tube would correlate the two measurements and bias the result. We therefore used an independent SiPM detector as the coincidence reference detector. The reference detector consisted of a 3.72×3.62 $mm^2$ active area SiPM (Broadcom AFBR-S4N44P014M) coupled to a 2×2×3 $mm^3$ LYSO crystal (Crystal Photonics, Inc., CPI) with optical glue. Custom readout electronics of the same design as described in Section II-B provided separate energy and timing outputs. Pulse time was determined by a digital leading edge discrimination (DLED) algorithm applied to the sampled timing waveform, with the threshold crossing time obtained by linear interpolation between samples, and the threshold was varied to find the value that minimized the CTR.

The reference detector was calibrated independently. For this, two identical SiPM detectors were operated in coincidence over a range of bias voltages, giving a best CTR of 58 ps FWHM (Fig. 3(a)), similar to values reported by other groups using high frequency amplifiers [19], [20], [21]. The CTR stays at its best value over a wide range of DLED thresholds (Fig. 3(b)). During the HRPPD measurements the reference SiPM was one of these two units, operating at the same 46 V bias, chosen to keep the noise level low and thereby reduce false triggers, yielding a CTR of 60 ps. Dividing by $\sqrt{2}$, a single reference detector contributes $T_S$ = 42 ps. The HRPPD detector's contribution is then

$$T_H = \sqrt{\mathrm{CTR}^2 - T_S^2} \quad (1)$$

where CTR is the measured CTR between the HRPPD and the reference detector. Two identical HRPPD detectors would give $\sqrt{2}T_H$. The sampled energy waveform was temporally integrated to estimate the deposited energy, and all events within the 511 keV photopeak were used for the coincidence timing analysis.

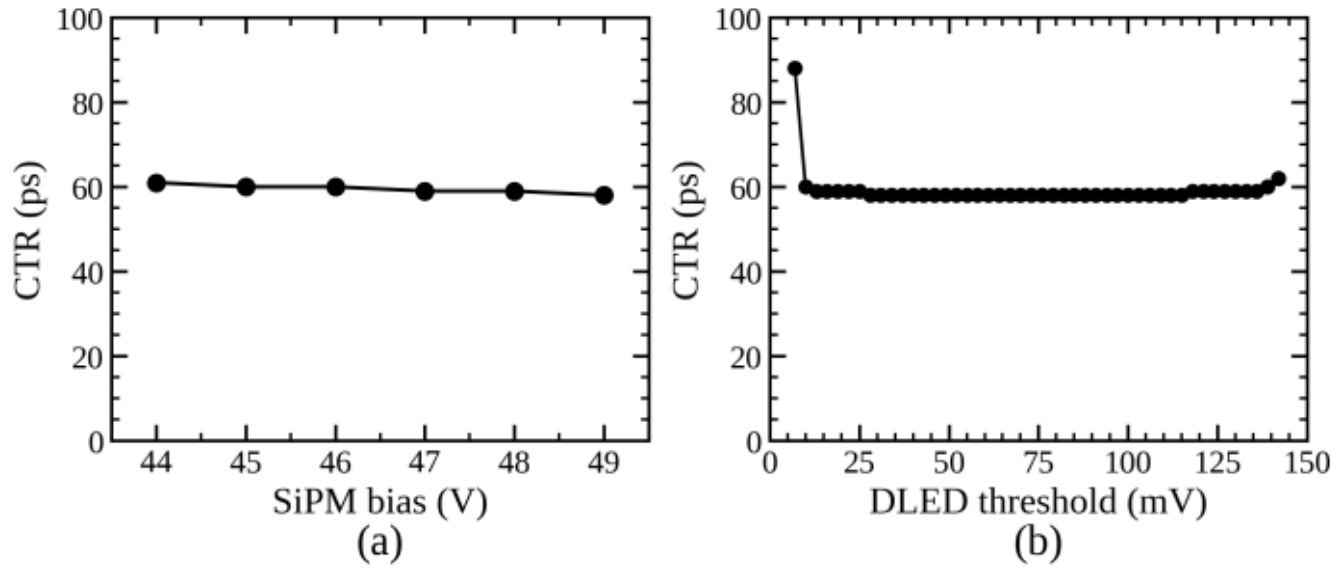

**Fig. 3.** Reference SiPM detector. (a) CTR versus bias voltage for two identical detectors, each a 2×2×3 $mm^3$ LYSO crystal on a 4×4 $mm^2$ SiPM. The best value is 58 ps FWHM. (b) CTR versus DLED threshold at 46 V, which is stable over a wide range.

## D. Measurements and Analysis

PET imaging generally requires a large area position sensitive detector with high timing resolution and moderate energy resolution. In this work we coupled the widely used LYSO scintillator to an HRPPD and measured the coincidence timing, the uniformity of timing and gain across pads, the 511 keV energy resolution, and the identification of individual crystals in an array, all at room temperature using a $^{22}$Na point source.

For the timing characterization we first established the working point of the HRPPD for best timing. A 2×2×3 $mm^3$ LYSO crystal (CPI), polished on all sides and wrapped with Teflon, was coupled to the HRPPD with silicone grease and centered on one pad of the lower left 2×2 subarray. The setup is shown in Fig. 4. The energy and timing outputs of the HRPPD and of the reference SiPM detector, each from its own readout electronics, were fed to the four channels of the oscilloscope, which recorded the waveforms of a coincidence whenever both energy channels exceeded a threshold set in the Compton region (Fig. 4). Approximately 20,000 such coincidences were acquired in each run. $V_{MCP}$ was first scanned from 600 to 655 V per MCP in 5 V steps at a fixed 200 V photocathode gap voltage ($V_{PG}$). $V_{PG}$ was then scanned separately with $V_{MCP}$ fixed at the optimal value found in the first scan. The 511 keV photopeak amplitude and the CTR were recorded at each setting, with DLED applied to the sampled timing waveform and its threshold varied to obtain the best CTR. The CTR of a clinical 3×3×20 $mm^3$ LYSO crystal (CPI) against the reference detector was then measured in the same way.

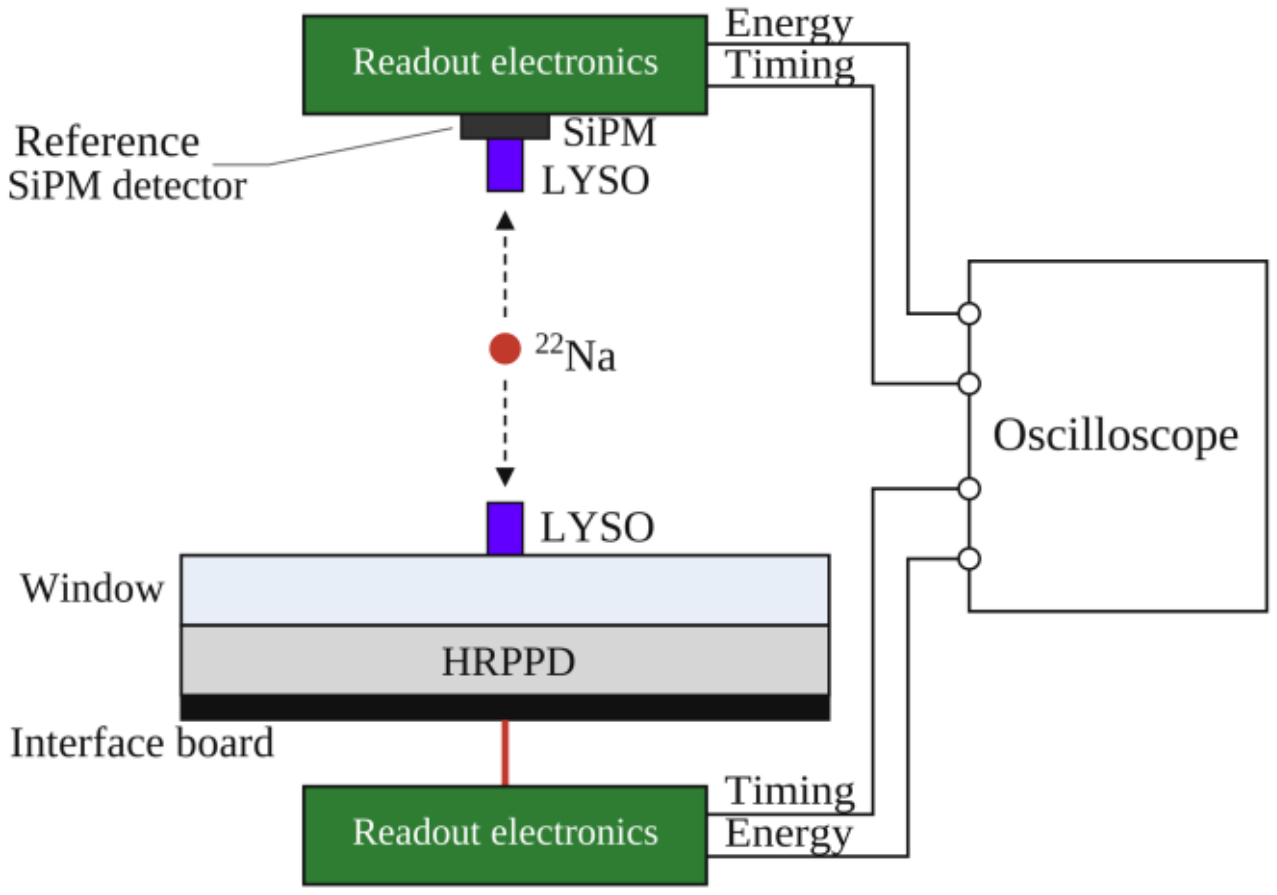

**Fig. 4.** Coincidence measurement setup. The LYSO crystal is coupled to one pad on the HRPPD window, and the reference SiPM detector with its readout electronics is mounted above. The $^{22}$Na point source is placed between the two detectors, and the energy and timing outputs of the readout electronics of each detector go to the oscilloscope.

The position dependence was then tested by centering the same crystal in turn on each of the four pads of the subarray at the working point found above, repeating the coincidence measurement and recording the CTR and the 511 keV photopeak position, which assesses timing and gain uniformity together.

The 511 keV energy resolution was measured in a dedicated run. Because the 5 mm entrance window distributes the scintillation light over several pads, a single pad collects only a portion of it. With the four channel DRS4 used in this study, the greatest and most equally shared fraction of the scintillation light is obtained by centering the crystal on the junction of the four pads that are read out. A 3×3×10 $mm^3$ crystal (CPI) was therefore placed in that position, shown in Fig. 5(a), and the charge of pads 9, 10, 13, and 14 was summed. Direct summing is conditional on the uniformity measurement above showing the gains of those four pads to be closely matched. No amplifier or shaping was applied in the readout electronics, so the deposited energy was taken as the integral of each waveform over a window running from 10 ns before the leading edge to 100 ns after it.

Flood maps of a 4×4 array of 3×3×10 mm³ crystals (JT Crystal Technology, JT) were acquired last, with the array placed at the center of the quadrant covered by the same four pads, as shown in Fig. 5(b). These pads were connected to the DRS4 board directly, with no front end electronics in between, and the map was formed by applying the center of gravity method to the integrated pad outputs. The centroid from this method is dimensionless. Each axis was converted to millimeters by fitting the four peak positions of the corresponding projection against the known 3 mm crystal pitch.

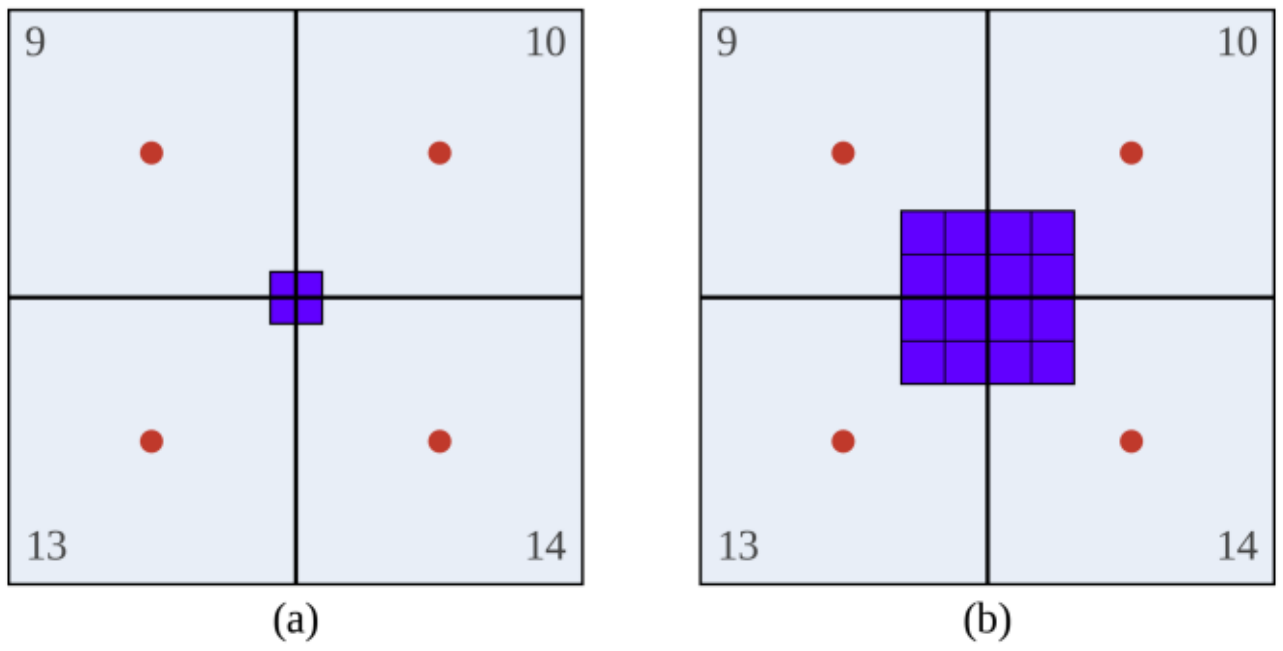

**Fig. 5.** Measurement configurations on the lower-left 2×2 subarray, pads 9, 10, 13, and 14. (a) Energy measurement with a 3×3×10 mm³ LYSO crystal centered on the junction of the four pads. (b) The 4×4 array of 3×3×10 mm³ crystals centered on the same quadrant, used for the flood map. Crystal footprints are not to scale.

CTR values are the FWHM of a Gaussian fit to the coincidence time difference distribution, restricted to photopeak events in both detectors. The dominant uncertainty is run to run repeatability. Between runs the crystal is positioned by hand and thus not at exactly the same point on the window. The crystals are wrapped in Teflon and are not re-wrapped between runs, but the grease is re-applied each time, so its wetting changes slightly and with it the fraction of light the HRPPD collects. These effects are individually small, but they act directly and collectively on the measured amplitude and timing and are included in the error budget, together with a smaller contribution from voltage drift of the bias supplies. From repeated measurements using identical configurations, the resulting scatter amounts to ~4 ps in CTR and ~5% in photopeak amplitude. The statistical uncertainty of the Gaussian fit is smaller, below 1.5 ps at the event counts used here. We therefore take ≈ 4 ps for CTR and about 5% for the photopeak amplitude. For the energy resolution, the fit uncertainty is about 0.1 percentage point, and the larger uncertainty comes from crystal placement and the choice of fit window.

## III. Results

### *A. Operating Point and Coincidence Time Resolution*

Fig. 6 shows the 511 keV photopeak amplitude and CTR as a function of MCP voltage at a fixed 200 V photocathode gap, measured with the 2×2×3 mm³ crystal. The amplitude follows an approximately exponential trend, doubling roughly every 12.9 $V_{MCP}$ and spanning a factor of ~19 (9 to 169 mV) over the 600 to 655 V range. When $V_{MCP}$ was below 600 V, the signal was too small for stable triggering. The CTR improves from 141 ps at 600 V to 107 ps at 650 V, more slowly at the higher MCP voltages.

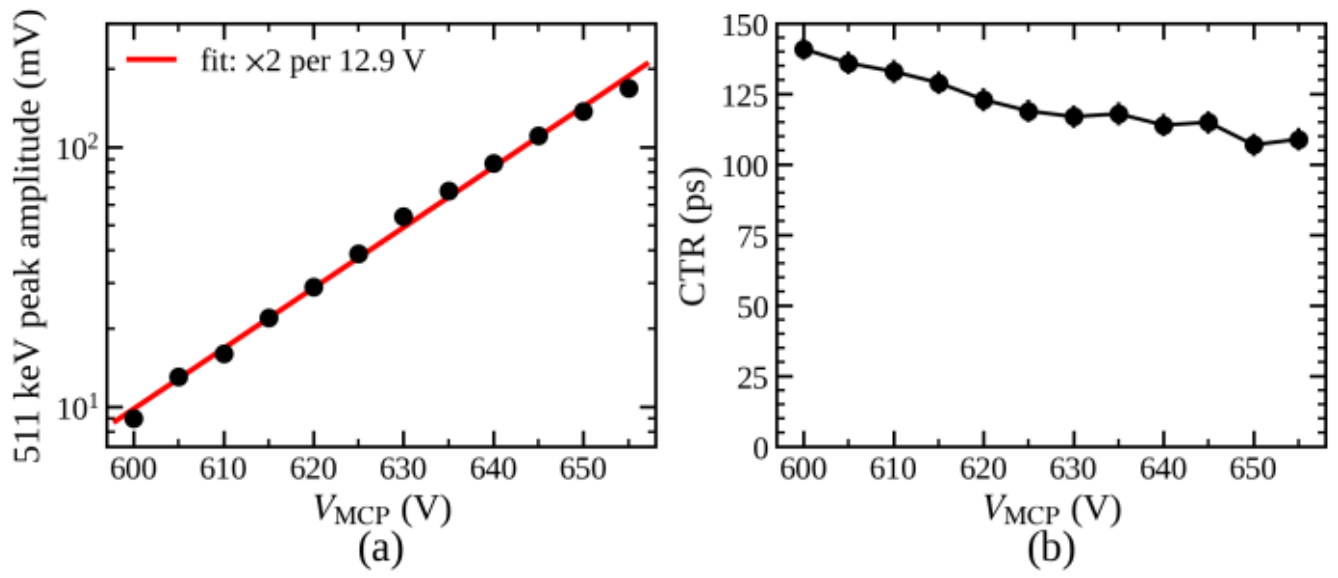

**Fig. 6.** MCP-voltage scan at a fixed 200 V photocathode gap, 2×2×3 mm³ crystal. (a) 511 keV photopeak amplitude with an exponential fit (doubling every 12.9 V). (b) CTR versus MCP voltage.

The photocathode gap scan at fixed MCP voltage of 650 V (Fig. 7) shows the same monotonic trend, with photopeak amplitude growing approximately linearly with gap voltage at a rate of ~0.48 mV/V. Over the range of 50 to 250 V photocathode gap voltages, the measured CTR falls from 137 ps to 108 ps, with almost all of the improvement occurring below 150 V and no significant gain up to the 250 V limit of the scan. We therefore fixed the operating point at $V_{MCP}$ = 650 V and $V_{PG}$ = 200 V, corresponding to electrode voltages of −1900, −1700, −1050, −850, and −200 V on HV #1 to HV #5, respectively, for everything that follows.

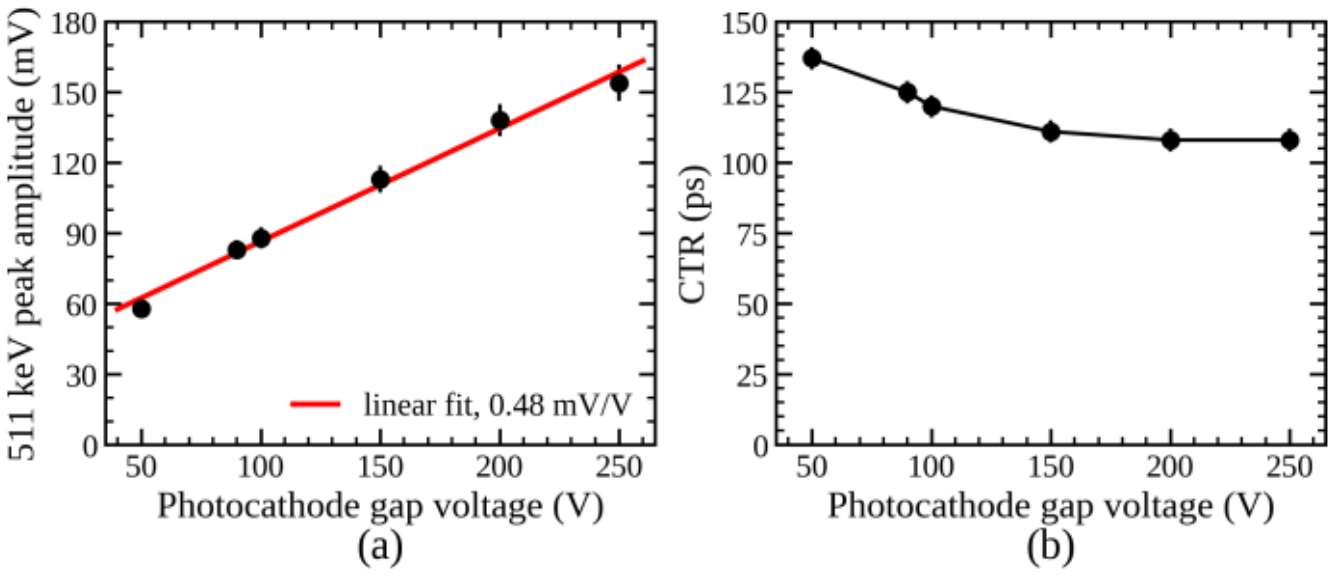

**Fig. 7.** Photocathode-voltage scan at the fixed MCP voltage. The gap voltage is the absolute photocathode voltage minus 1700 V. (a) 511 keV peak amplitude with a linear fit (0.48 mV/V). (b) Measured CTR against the SiPM reference detector.

At this operating point the best CTR we obtained with the 2×2×3 mm³ crystal was 107 ps FWHM. Subtracting the calibrated reference detector's contribution of 42 ps in quadrature leaves 98 ps for the HRPPD detector, which corresponds to 139 ps for two identical HRPPD detectors operating in coincidence. The same measurement was repeated using the clinical 3×3×20 mm³ LYSO crystal and yielded a CTR of 160 ps (Table I). Subtracting the calibrated reference detector's contribution reveals that each HRPPD detector contributes 154 ps, which corresponds to a CTR of 218 ps for two identical HRPPD detectors operating in coincidence.

TABLE I
MEASURED AND DECONVOLVED TIMING FOR THE BENCHMARK AND CLINICAL CRYSTALS

| Crystal (mm) | 511 keV peak (mV) | CTR (ps) | Single HRPPD (ps) | HRPPD pair (ps) |
|---|---|---|---|---|
| 2×2×3 | 140 | 107 | 98 | 139 |
| 3×3×20 | 98 | 160 | 154 | 218 |

Operating point 650 V/MCP and 200 V photocathode gap. The deconvolution uses $T_S$ = 42 ps. Uncertainty is approximately ±4 ps on CTR.

Fig. 8(a) and (b) show the time difference spectra at the best DLED threshold and Fig. 8(c) and (d) show the dependence of the CTR on the HRPPD threshold for two different sizes of LYSO crystals. The time difference is the reference detector time minus the HRPPD detector time, and the reference detector threshold is fixed at 50 mV. Both spectra have a tail on the late side of the HRPPD signal, with 2.4% (2×2×3 mm³) and 3.0% (3×3×20 mm³) of the events beyond 2σ. Since the reference detector alone gives a symmetric spectrum, the tail originates in the HRPPD channel and is larger for the longer crystal. The threshold crossing is determined by the earliest photons to arrive. So a wider spread of optical path lengths and a smaller photoelectron count could result in crossing later, with greater effect in longer crystals. Its impact is small in the present measurement, but worthy of consideration once the coincidence between two HRPPD detectors becomes possible. The CTR reaches its best value at a low HRPPD DLED threshold and remains flat over most of the scan. For the 2×2×3 mm³ crystal the CTR is 107 ps at 10 mV and stays below 112 ps up to 150 mV. For the 3×3×20 mm³ crystal the CTR is 160 ps at 15 mV and stays below 168 ps up to 150 mV. Such a wide plateau makes a fixed-threshold leading-edge discriminator, with the TDC electronics common in PET systems, a viable option for the readout.

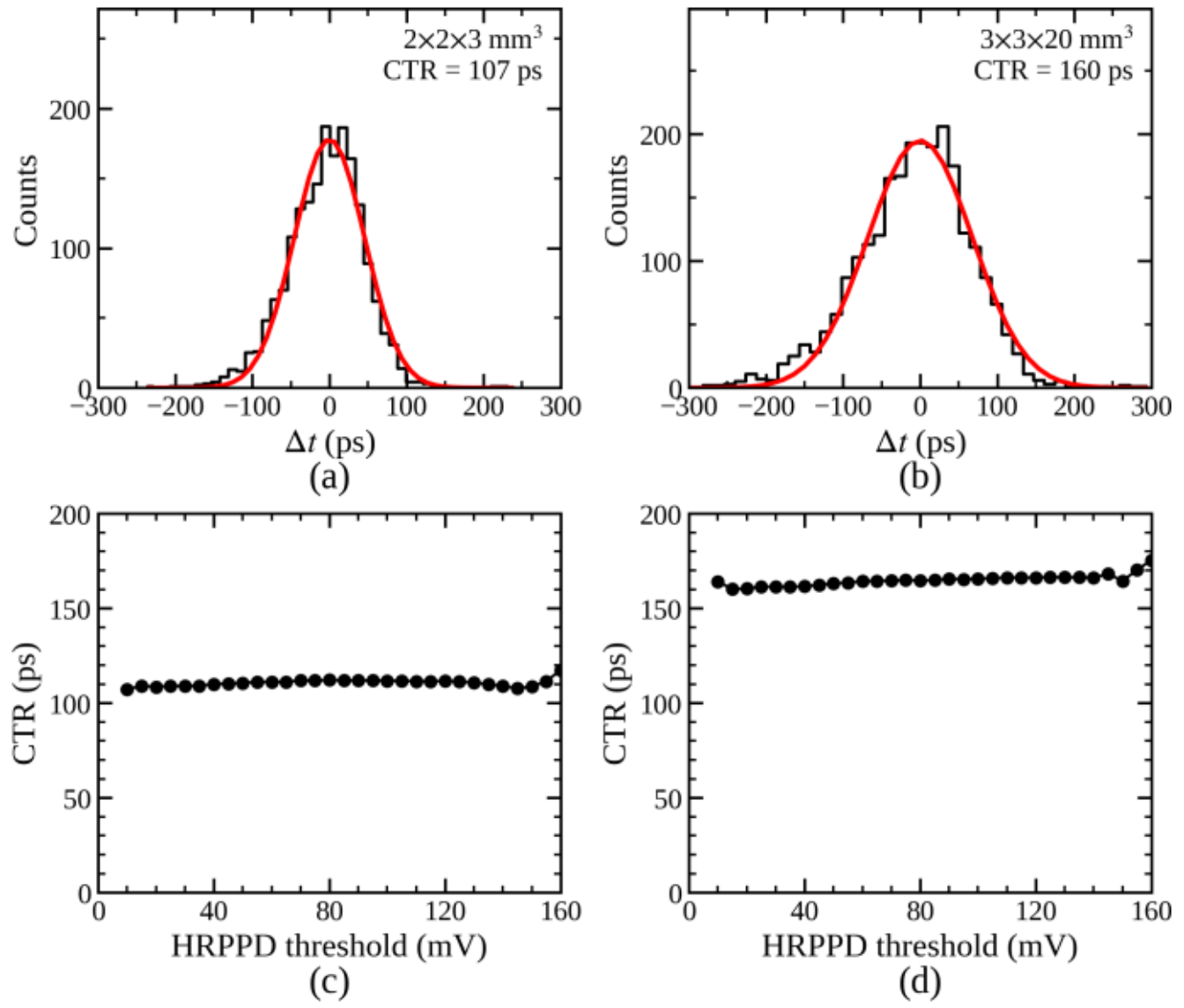

**Fig. 8.** Coincidence timing against the SiPM reference detector with reference detector threshold of 50 mV. (a), (b) Time difference spectra at the best HRPPD threshold, plotted relative to the fitted mean, with the Gaussian fit in red. (c), (d) CTR against the HRPPD DLED threshold. Left column, 2×2×3 mm³ crystal. Right column, 3×3×20 mm³ crystal.

### *B. Pad-to-Pad Uniformity*

The 2×2×3 mm³ crystal was repositioned at the center of each pad in the lower left 2×2 subarray (Table II). Measurements were then made of photopeak amplitude, which ranged from 132 to 150 mV, and CTR, which ranged from 107 to 117 ps. Given the ±4 ps repeatability of the CTR measurements, none of the pad-to-pad CTR differences reached 2σ, and the photopeak amplitude spread (±6% about mean) is comparable to the ±5% measurement precision. Because the manual placement of the LYSO crystal, its Teflon wrapping and grease wetting, and the stability of bias all contribute at comparable levels, these measurements can support only the broader conclusion that the four pads behave alike, with no systematic differences among them. The four pads sample the corner, edge, and interior positions of the array and share an identical readout chain. Mapping the full detector at this precision is limited by acquisition throughput, since the four oscilloscope channels are shared between the reference SiPM detector and the HRPPD energy and timing outputs so that only one pad is read at a time. This is left to future work with dedicated instrumentation.

TABLE II
PAD-TO-PAD UNIFORMITY OVER THE LOWER LEFT 2×2 SUBARRAY

| Pad | 511 keV peak (mV) | CTR (ps) |
|---|---|---|
| 9 | 132 | 117 |
| 10 | 138 | 107 |
| 13 | 149 | 110 |
| 14 | 150 | 115 |

Uncertainties are about ±5% on the peak amplitude and ±4 ps on CTR.

### *C. Energy Resolution and Linearity*

The deposited energy is represented by the waveform integral in mV·ns. Into the 50 Ω input, 1 mV·ns corresponds to 20 fC of collected charge. As shown in Fig. 9, the 511 keV photopeak positions of the four pads range from 2781 to 2932 mV·ns, within ±2.8% of their mean. Each photopeak was fitted with a Gaussian without a background term, over a window iterated to ±1.5σ about the fitted center, and resolution is determined by the fit's FWHM divided by its centroid. The individual channels resolve the 511 keV photopeak at 16.2% to 16.8% FWHM, while the four-pad sum gives 14.6% FWHM (Fig. 9 and Table III). The $^{22}Na$ spectra also resolve the 1275 keV line as a linearity check. The summed 1275 keV peak is at 28952 mV·ns against 11526 mV·ns at 511 keV, a ratio of 2.512 against the expected 2.494, so the summed response is linear to better than 1% over this range.

The resolution as it stands is sufficient to define the 511 keV window used for coincidence selection. Two issues, however, impact these measurements. No per-channel gain correction was applied to the sum, and with four of the sixteen pads digitized at a time the sum captures only the light falling within one quadrant. So the result depends upon where the crystal is placed relative to the pads. With a dedicated sixteen-channel readout this issue can be readily resolved, and all the light would be collected and each channel individually corrected, which would improve the resolution and make the position dependence measurable.

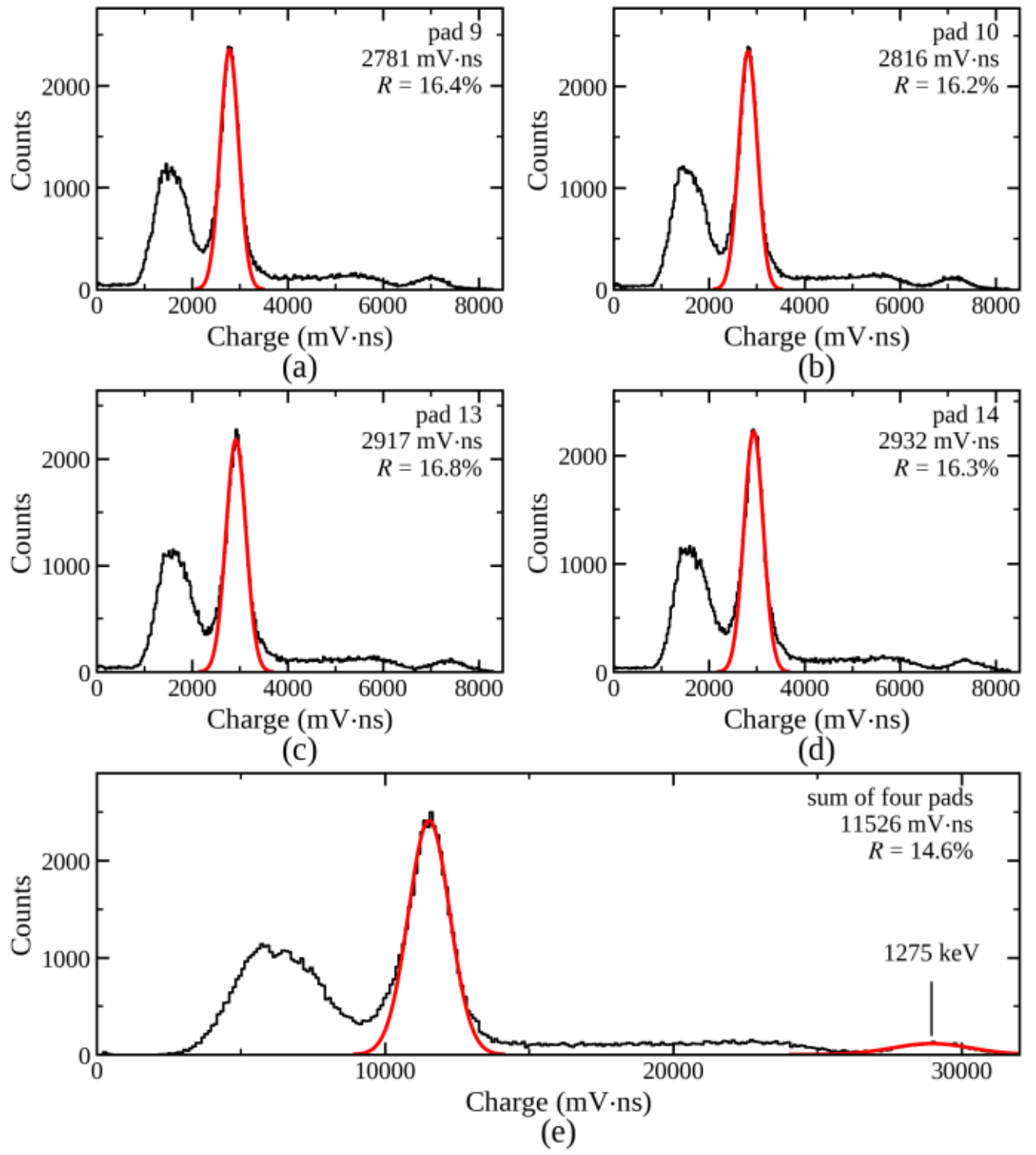

**Fig. 9.** Charge spectra of the four pads, (a) to (d), and of their sum, (e), for the 3×3×10 mm³ crystal in the configuration of Fig. 5(a). The red curves are Gaussian fits, to the 511 keV photopeak in every panel and to the 1275 keV line of $^{22}$Na in (e). The four pads collect nearly equal charge, and the sum gives an energy resolution ($R$) of 14.6% FWHM at 511 keV.

TABLE III
PER-PAD AND 4 PAD SUMMED ENERGY RESOLUTION WITH THE CRYSTAL ON THE PAD JUNCTION

| Channel | 511 keV peak | Energy resolution | 1275 keV peak |
|---|---|---|---|
| Ch9 | 2781 | 16.4% | — |
| Ch10 | 2816 | 16.2% | — |
| Ch13 | 2917 | 16.8% | — |
| Ch14 | 2932 | 16.3% | — |
| Sum | 11526 | 14.6% | 28952 |

Charges are in mV·ns. The 1275 keV line of the $^{22}$Na source serves as a linearity check, and it is fitted only on the summed spectrum.

### *D. Crystal-Array Identification*

Fig. 10 shows a flood map taken with the DRS4 readout and center-of-gravity positioning on the 2×2 pads spanning the array. As the figure illustrates, each of the 16 elements of the 4×4 array of 3×3×10 mm³ crystals is clearly resolved. One-dimensional projections of the crystal flood map peaks reveal FWHMs of 1.11 mm in X and 1.16 mm in Y, since the centroid averages over many photoelectrons and its event-to-event spread is much smaller than the crystal's cross section. Peak-to-valley ratios average 4.9 in X and 4.7 in Y. This indicates that, despite its coarse 25 mm pad pitch, the HRPPD can provide sufficient spatial resolution for clinical PET imaging. To our knowledge this is the first crystal-array identification demonstrated on a LAPPD or HRPPD.

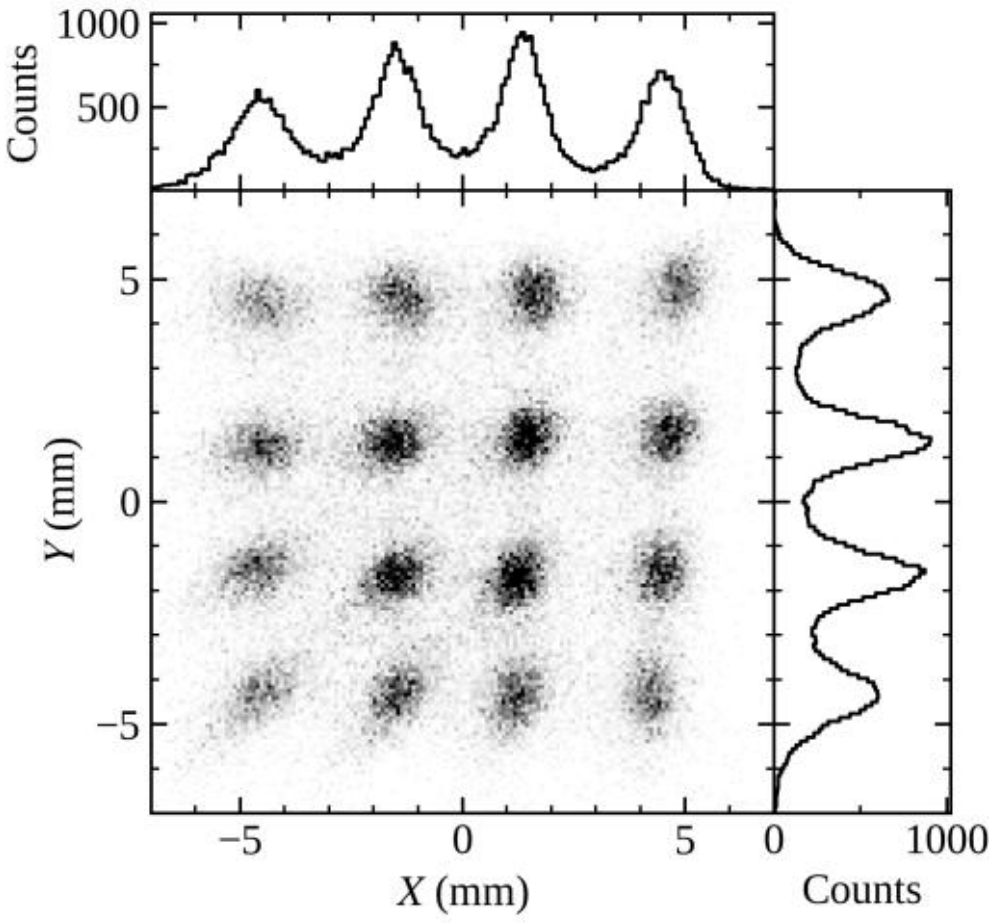

**Fig. 10.** Flood map of a 4×4 array of 3×3×10 mm³ LYSO crystals obtained by center-of-gravity positioning over the 2×2 pads spanning the array. All 16 crystals are fully resolved despite the 25 mm pad pitch.

## IV. Discussion

The intrinsic performance of MCP-PMTs in general, and their SPTRs in particular, has been demonstrated repeatedly by independent groups on tubes from numerous manufacturers [1], [18], [22], [23]. What has been missing is a large-area tube coupled to a PET scintillator that performs at the level its single-photon response suggests. The results reported here narrow that gap. For the clinical 3×3×20 mm³ geometry the equivalent CTR for two identical HRPPD detectors is 218 ps, with a 511 keV energy resolution of 14.6% from the summed pads and the 1275 keV peak within 1% of its proportional position. The gain and timing are uniform across the pads within measurement precision, and the individual 3 mm elements of a 4×4 array are spatially resolved. These metrics place the detector close to those of state-of-the-art clinical TOF-PET systems [24], [25], and none of them is far from what a clinical module would need. The present design nonetheless leaves headroom for optimization.

Improved CTR in this study is the main advantage over the earlier large-area MCP-PMT work. With a 2×2×3 mm³ LYSO crystal on a single pad, the best measured CTR against the reference SiPM is 107 ps, which deconvolves to 98 ps for the single HRPPD channel and 139 ps for two identical HRPPD detectors operating in coincidence. For a clinical 3×3×20 mm³ LYSO crystal at the same operating point, the equivalent CTR for two identical HRPPD detectors is 218 ps. The HRPPD used here is the capacitively coupled version (CC-HRPPD), in which the interior anode couples through the ceramic base to the pads of an external readout board. Optically, the 5 mm fused-silica window separates the crystal from the photocathode, so the light spreads laterally before reaching the photocathode and photons emitted at the same instant travel different path lengths. For a crystal centered on a pad, a geometric estimate puts about 85% of the light entering the window within that 25 mm pad. Among the photons the pad collects, the path through the window ranges from 5.0 mm along the normal to 18.4 mm at the pad corner, so their arrival times differ by up to 65 ps. Electrically, each channel of the present board reads out a 25 mm pad, and the signal induced at different points of the pad travels different distances to the central output electrode, which introduces a position-dependent delay and hence additional timing jitter. Both effects broaden the arrival-time distribution of the photoelectrons that form the leading edge of the signal, which degrades the timing. To mitigate these effects, an HRPPD with a thinner window would reduce both the lateral spread and the time spread. And because the readout board is external to the sealed tube, it can be replaced by a board with smaller pads without changing the device. Smaller pads narrow the spread of both optical and electrical delays within one channel, and a separate time calibration of each pad removes the remaining offsets between channels. The direct-coupled version (DC-HRPPD), developed for Ring Imaging Cherenkov (RICH) detectors at the Electron-Ion Collider (EIC), provides finer segmentation by construction. It shares the same window thickness, photocathode and MCP stack [18], but its 1024 square anode pixels (3 mm side, 3.25 mm pitch) are routed through the multilayer ceramic anode to readout connectors on the air side of the device. The smaller pixels offer more flexibility for PET readout, since they can be read out individually or combined into channels of the size a given design requires. Taken together, modest structural changes of this kind are expected to further improve the CTR.

Beyond timing, the HRPPD brings a set of advantages that are intrinsic to the MCP, compared with the SiPM, the workhorse photodetector of PET today. The results in this work came from 25 mm pads, which is about as coarse and as simple a readout as the device permits. Collecting the signal over such a large area with a single channel costs little because the capacitance is low. In the CC-HRPPD the external pads are separated from the interior anode by a 2 mm thick ceramic plate ($\varepsilon_r \approx 10$). A parallel-plate estimate gives about 0.044 pF/mm$^2$, or about 28 pF for a single $25 \times 25$ mm$^2$ pad, and a direct measurement including stray capacitance gives an upper limit of 142 pF. Even this upper limit is below the several hundred pF of a single typical SiPM [26]. An area of about 100 cm$^2$ for HRPPD, or about 370 cm$^2$ for a LAPPD, can therefore be covered at a low channel count. The current HRPPD pores are 10 μm across and smaller ones are available, and each pore is an independent gain element. In addition to the dynamic range that follows from this, it means the position resolution of a module is defined by the granularity of the anode plate rather than by the intrinsic characteristics of the sensor, which leaves room for finer position sampling with a specific segmentation pattern. The low dark count rate makes Cherenkov-based PET workable, since those events produce only a handful of photons. Precise timing from simple electronics, finer spatial resolution, and a dynamic range wide enough for theranostic imaging all widen what an MCP-based PET detector could be.

The device is also expected to be less demanding to operate, requiring little of the bias compensation or active cooling that SiPM-based detectors typically rely on to keep their gain and dark count rate stable. This would simplify both the electronics and the thermal design of a system. The radiation tolerance of a vacuum device becomes relevant in theranostic imaging, where the activities can be higher than in diagnostic PET and the radiation field is mixed. In life testing by Agarwala et al. [17], delivering $6\times10^{14}$ photons/cm$^2$ from a light source reduced the quantum efficiency by only 1%, indicating minimal photocathode degradation at this exposure.

Limited by the channel count of the amplifier and the data acquisition system, this study presents a CTR measurement of single HRPPD channels against a reference SiPM, while the position and energy resolution cover four pads in one quadrant without dedicated amplification, shaping or per-channel calibration. The next steps that follow from these will be the development of a detector module, including dedicated multi-channel readout electronics with full timing and position decoding, followed by a full evaluation covering timing resolution, position resolution, count rate and so on. On the device side, an HRPPD with finer pad segmentation and a thinner window should improve the timing and the position resolution simultaneously. Simulation, modelling and HRPPD testing that quantitatively characterize these contributions are currently underway.

## V. Conclusion

We have characterized a $104\times104$ mm$^2$ HRPPD MCP-PMT with $4\times4$ pads on a 25 mm pitch, coupled to LYSO scintillation crystals for TOF-PET. With a custom high-frequency readout and an optimized operating point, the detector achieves a CTR of 107 ps FWHM against a fast SiPM reference detector using a $2\times2\times3$ mm$^3$ LYSO crystal, and a CTR of 160 ps using a clinical $3\times3\times20$ mm$^3$ LYSO crystal. Deconvolving the reference detector's contribution gives an equivalent CTR of 139 ps for two identical HRPPD detectors with $2\times2\times3$ mm$^3$ crystals and 218 ps for two identical HRPPD detectors with $3\times3\times20$ mm$^3$ crystals. Summing the four pads under the crystal gives a 511 keV energy resolution of 14.6% FWHM, with the 1275 keV peak of $^{22}$Na within 1% of its proportional position. The individual crystals of a $3\times3\times10$ mm$^3$ array are clearly resolved despite the 25 mm pad pitch, and timing and gain are uniform across pads within measurement precision. These PET-related results show that coupling a large-area MCP-PMT to LYSO is potentially a viable route for clinical TOF-PET system design. They indicate that the previous gap between the single-photon timing of the device and the CTR obtained with a scintillator can be addressed by dedicated external electronics, and motivate structural refinements of the device for further CTR improvement.

## Acknowledgment

M. A. Popecki, A. V. Lyashenko, S. Cwik, and M. J. Minot are employed by Incom Inc., which manufactures the HRPPD studied in this work. The remaining authors declare no competing financial interests or personal relationships that could have influenced the work reported in this article.